\documentclass[twocolumn,prl, letterpaper,preprintnumbers,amsmath,amssymb]{revtex4}

\usepackage{graphicx}% Include figure files 
\usepackage{subfigure} 
\usepackage{dcolumn}% Align table columns on decimal point 
\usepackage{array} 
\usepackage{bm}% bold math 
\usepackage[hang]{footmisc} 

\usepackage[usenames,dvipsnames]{color}

\begin{document}

\title{\bf{High-throughput first principles search for new ferroelectrics} \\[11pt] } 
\author{Kevin F. Garrity}
\email{kevin.garrity@nist.gov}
\affiliation{%
Material Measurement Laboratory, National Institute of Standards and Technology, Gaithersburg MD, 20899
}%

\date{\today}% It is always \today, today, % but any date may be explicitly specified 

\begin{abstract}
We use a combination of symmetry analysis and high-throughput density
functional theory calculations to search for new ferroelectric
materials.  We use two search strategies to identify candidate
materials. In the first strategy, we start with non-polar materials
and look for unrecognized energy-lowering polar distortions. In the
second strategy, we consider polar materials and look for related
higher symmetry structures.  In both cases, if we find new structures with
the correct symmetries that are also close in energy to experimentally known structures, then the material is likely to be
switchable in an external electric field, making it a candidate
ferroelectric. We find sixteen candidate materials, with variety of
properties that are rare in typical ferroelectrics, including large polarization,
hyperferroelectricity, antiferroelectricity, and multiferroism. 
\end{abstract}

\maketitle 

Ferroelectrics, which are materials that have a ground state polar
phase that can be switched to a symmetry-equivalent structure by the
application of an external electric field, have been studied and used
in applications for many years. Much of the attention on
ferroelectrics has focused on prototypical examples from the
perovskite oxides, like PbTiO$_3$ and BiFeO$_3$. However, in recent
years, there has been a renewed theoretical and experimental interest
in discovering and understanding the properties of new ferroelectric
materials\cite{abcferro,abcantiferro,halfheusler,hyperferro,hybridimproper,ymno3_origin,ymno3,dionjacob,zro2,
  eutio3,switch_hybrid_improper,eutio3_expt,ymno3_domainwalls, srmno3}.  These new materials have shown a variety of new or rare
behaviors, including hybrid improper ferroelectricity,
hyperferroelectricity, topological defects/domain walls, etc.  In addition,
there is interest in and need for ferroelectrics with improved
functionality for various applications, including
magnetoelectrics, Pb-free piezoelectrics, room temperature
multiferroics, solar energy converters, silicon compatible
ferroelectrics, ferroelectric catalysts, etc.\cite{garrity_fe_surface,ferro_surf_review,shiftcurrent,silicon_oxide,hfo2_devices,few_multiferroics}

High-throughput first principles density functional theory (DFT)
calculations have been used increasingly in recent years as a tool for
materials discovery and
screening\cite{compmatsci,datamine,wolverton,materialsproject,aflowlib,oqmd}. DFT
calculations using semilocal functionals are generally reliable tools
for computing ground state properties and the energy differences
between closely related phases, which is the primary screening tool
needed to identify ferroelectrics. The main obstacle to finding new
ferroelectric materials computationally is identifying the relevant
polar and non-polar structures to consider.  Any insulator known experimentally to
have both a polar and a non-polar phase has likely already been
recognized as a potential ferroelectric, necessitating a search for
new structures of known compounds. However, systematically searching all insulating
compounds for possible new phases is too computationally demanding to
attempt systematically.

In this work, we employ two strategies to identify new ferroelectrics, 
which we apply to a much larger range of materials than
related previous searches\cite{abcferro,joedatabase,joe_icsd,eutio3,srmno3,abrahams1,abrahams2,abrahams3,
  acentric,noncentro_oxides, pb_free}.  In the first strategy, aimed
at finding proper ferroelectrics, we start with a list of
experimentally known non-polar transition metal oxides, nitrides, and
sulfides, and we perform a $\Gamma$-point phonon calculation, looking
for materials with unstable polar modes. In order to reduce the
computational cost of this step, we reuse phonon calculations from
our earlier study of transition metal
thermoelectricity\cite{mythermoelectrics}, and we supplement this list
with additional calculations of materials with small unit cells. After
identifying materials with unstable polar distortions, we proceed
to look for the ground state structure, including possible non-polar distortions. If the ground state is polar,
then the material is a candidate ferroelectric.

In the second strategy, we begin instead with a list of materials
already known to be polar but not known to be switchable in an
external electric field. We attempt to identify switchable materials
by using a symmetry search algorithm\cite{pymatgen} and adjusting the
tolerance factor systematically to allow the algorithm to identify
possible higher symmetry structures that are related to the polar
ground state. We then calculate the energy of these new structures,
and if they are close in energy to the polar structure, the materials
are likely to be switchable\cite{abcferro}. An advantage to both of
our search strategies is that they are not limited to generating previously observed
structure types, allowing us to consider new ferroelectric mechanisms.

In the remainder of this letter, we will detail our search strategy,
identify candidate ferroelectrics, and discuss some of their
interesting properties. We find materials with unusual chemistry for
ferroelectrics, as well as potentially useful properties that include strong polarization, magnetism,
hyperferroelectricity, and antiferroelectricity. In addition, we
rediscover and identify the missing structures of two materials that
were previously studied as ferroelectrics but have not been fully
characterized. Full structural details our of candidate materials can
be found in the supplementary materials.

We perform first principles DFT calculations\cite{hk,ks} with a
plane-wave basis set as implemented in QUANTUM ESPRESSO\cite{QE} and
using the GBRV high-throughput ultrasoft pseudopotential
library\cite{ultrasoft,gbrv}. We use a plane wave cutoff of 40 Ryd for band
structure calculations and 45-50 Ryd for phonon calculations\cite{phonon_convergence}. For
Brillouin zone integration, we use a $\Gamma$-centered grid with a
density of 1500 k-points per atom.

We use the PBEsol exchange-correlation functional\cite{pbesol}, which
provides more accurate lattice constants and phonon frequencies than
other generalized gradient approximation 
functionals\endnote{Except for CuBiW$_2$O$_8$, which uses LDA\cite{lda1}, see
  later in text}. We perform phonon calculations using DFT
perturbation theory\cite{dft-pt}, and polarization calculations with
the Berry phase method\cite{modern_polarization}.  We use
PYMATGEN\cite{pymatgen} to manipulate files from the Inorganic Crystal
Structure Database (ICSD) to setup the initial structures for
relaxation. We relax each structure several times to ensure
consistency between the basis set and the final structure, with a
force tolerance of 0.001 Ry/Bohr, an energy tolerance of $1 \times
10^{-4}$ Ry, and a stress tolerance of 0.5 Kbar. For phonon
calculations, we decrease the force tolerance to $5 \times 10^{-5}$
Ry/Bohr.

To identify ferroelectrics, we look for materials that are a)
insulating, b) have a polar ground state, and c) have a higher
symmetry reference structure that is close in energy to the polar
structure, which we take as an indication the material is likely to be
switchable in an external electric field. Some examples of this energy
difference for known ferroelectrics include PbTiO$_3$ (20 meV/atom),
Ca$_3$Ti$_2$O$_7$ (40 meV/atom), ZrO$_2$ (16
meV/atom)\cite{zro2,hybridimproper}. Of course, in real
ferroelectrics, the switching proceeds via a combination of domain
wall nucleation and motion, and does not correspond directly to the
first principles energy difference; however, a small energy difference
has proven to be a useful indicator of whether a material is likely to
be switchable.

\begin{table} 
\begin{center} 
\begin{ruledtabular} 
\begin{tabular}{lcccccccc} 
Composition & High-sym.   & Polar  & $\Delta$E & Polariz. \\
 & space grp. & space grp. & (meV/atom)  & $\mu C/cm^2$          \\
\hline
SrNb$_6$O$_{16}$         & $Amm2$     &  $Cm$    &   5.5     &   7\\
NaNb$_6$O$_{15}$F        & $Amm2$     &  $Cm$    &   9.5     &   6\\
RbCa$_2$Nb$_3$O$_{10}$   & $P4/mmm$     &  $Pc$    &   46.6   &   23\\
BaBi$_2$Ta$_2$O$_9$      & $I4/mmm$    &  $Cmc2_1$    &   15.7   &   37\\
LiScAs$_2$O$_7$   &  $C2$    & $P1$     & 1.6   &   7\\
YSF   &  $P6_3/mmc$   & $P6_3mc$    &   11.1   &   1.2 \\
\hline
CuBiW$_2$O$_8$   &  $P\overline{1}$    & $P1$     &   0.3   &   6\\
PbGa$_2$O$_4$   &   $P\overline{6}2c$    & $Ama2$     &   108   &   23\\
PbAl$_2$O$_4$   &   $P\overline{6}2c$    & $Ama2$     &   75   &   18\\
LiV$_2$O$_5$   &   $Pmmn$    & $Pmn2_1$     &   0.4   &   9\\
NaVO$_2$F$_2$   &  $P2_1/m$    & $P2_1$     &   0.3   &   18\\
SbW$_2$O$_6$   &   $P2_1/c$    & $P2_1$     &   4.4   &   21 \\
V$_2$MoO$_8$   &  $Cmmm$    & $Cmm2$     &   84   &   108 \\
Zn$_2$BrN   &  $Pnma$    &  $Pna2_1$    &   1.3   &   1.4\\
Zn$_2$ClN   &  $Pnma$    &  $Pna2_1$    &   2.4   &   1.5\\
AlAgO$_2$   &  $Pnma$    &  $Pna2_1$    &   71     &   0.2 \\
\end{tabular} 
\end{ruledtabular} 
\caption{Data on candidate ferroelectrics, found by starting from high
  symmetry (top) and low symmetry (bottom) structures. The second and
  third columns are space groups of the high and low symmetry
  structures, column four is the energy difference in meV/atom, and
  the last column is the polarization in $\mu C/cm^2$. }
\label{tab:results} 
\end{center} 
\end{table} 

In our first search strategy, the initial screening step is to perform
a $\Gamma$-point phonon calculation, which we did for 267 transition
metal oxides, nitrides, and sulfides. We identified 90 compounds with
unstable modes at $\Gamma$, although some of these distortion were
non-polar.  After excluding known ferroelectrics and materials with
known non-polar distortions, we proceeded to search for the ground
state structure. This search consisted of first calculating the phonon
dispersion of the material, and then freezing in finite amounts of
each unstable phonon eigenvector, as well as pairs of eigenvectors,
into supercells that correspond to points in the Brillouin zone with
unstable modes. We then relax the resulting
structures\cite{srmno3,bifeo3,interface_structure_determination}. For
systems with many unstable modes, which could have ground states
consisting of complicated distortion patterns, we supplemented this
searching with a random search
strategy\cite{interface_structure_determination}. To perform a random
search, we take a supercell, freeze in small random distortions, and relax.  
For most materials, we found that after a set
of ten random relaxations in a given supercell, the same low energy
structures repeated several times, and we terminated the search.

After this search process, if a material has a polar ground state,
then the material is a candidate ferroelectric. In addition, if the
material has competing polar and antipolar phases, then the material is a
candidate antiferroelectric\cite{rabeantiferroelectricity,abcantiferro}. In the top rows of table
\ref{tab:results}, we present six candidate ferroelectrics discovered
with this strategy. We proceed to briefly discuss several of these structures.

First, we note that after doing our analysis, we discovered that
BaBi$_2$Ta$_2$O$_9$ and related materials with Sr and Ca have has previously been studied as a
ferroelectrics\cite{babitao}. However, only a pseudotetragonal twinned
version of the polar structure of BaBi$_2$Ta$_2$O$_9$ had been
determined, explaining why it did not show up in our database search as a
polar. This material has a layered structure that consists of
perovskite-like BaTaO$_3$ layers separated by BiO layers, and these
layers shift relative to each other in-plane, resulting in a
relatively large polarization. Re-identifying a known ferroelectric
gives us confidence in the utility of our computational methodology.

\begin{figure}
\includegraphics[width=3.2in]{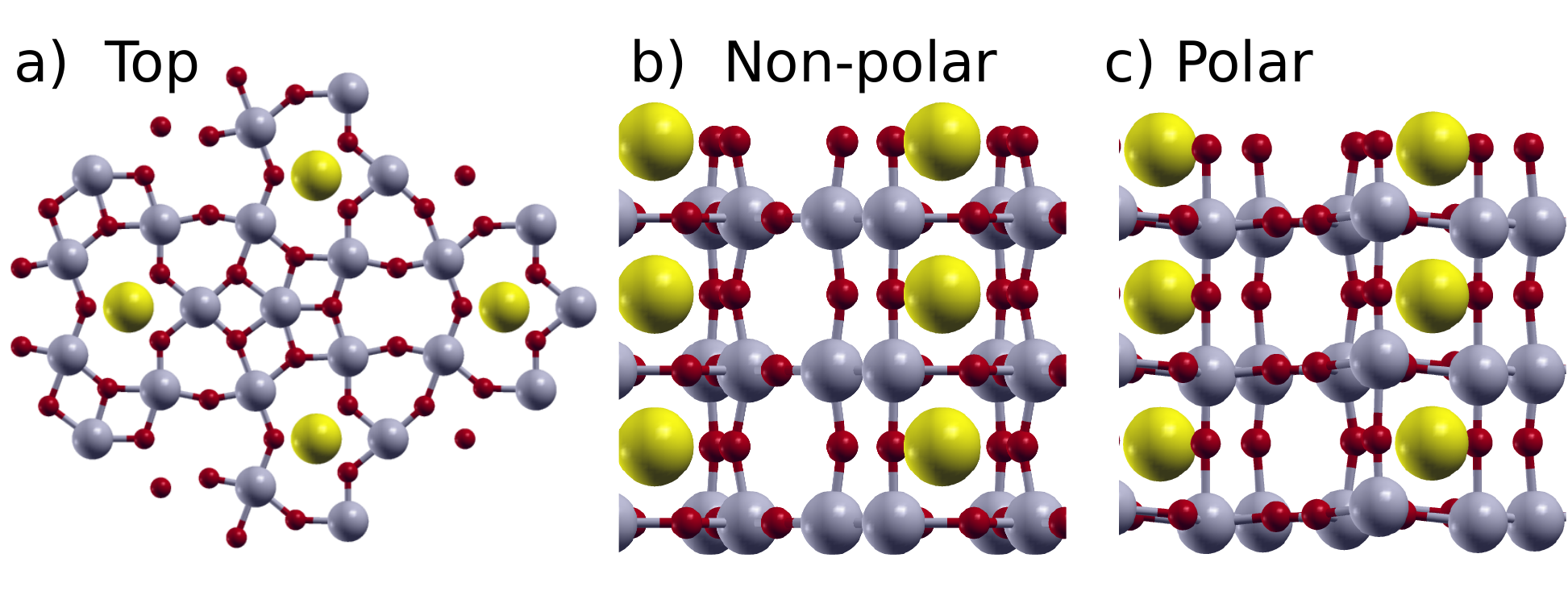}% Here is how to import EPS art
\caption{\label{fig:srnbo} Atomic positions of SrNb$_6$O$_{16}$. a)
  Top view, b) side view of non-polar structure, and c) side view of
  polar structure. The Sr are the large yellow atoms, the Nb are the 
medium gray atoms, and the O are the small red atoms. 
  direction.}
\end{figure}

\begin{table} 
\begin{center} 
\begin{ruledtabular} 
\begin{tabular}{ccccccccc} 
Mode   &  LO   & LO  & TO & TO \\
Number & Freq. $cm^{-1}$ & Mode $Z^*$ & Freq. $cm^{-1}$ & Mode $Z^*$ \\
\hline
1 & 176$i$& 21.8 & 139$i$ & 0.23 \\
2 & 138$i$& 5.9  & 128$i$ & 0\\
3 & 128$i$& 0 &  85$i$ & 1.2 \\
4 & 44$i$& 10.2 & 29$i$ & 0.32 \\
\end{tabular} 
\end{ruledtabular} 
\caption{Unstable LO and TO phonon frequencies and mode effective
  charges in SrNb$_6$O$_{16}$ at $\Gamma$. The first two columns are
  the frequency in $cm^{-1}$ and the mode Born effective charge of the
  LO modes, the second two columns are the same but for the TO
  modes. There is one anti-polar mode with zero mode effective charge
  that is unaffected by electrostatic boundary conditions.}
\label{tab:zeff} 
\end{center} 
\end{table}

SrNb$_6$O$_{16}$\cite{srnbo}, shown in Fig.~\ref{fig:srnbo}, and the closely related NaNb$_6$O$_{15}$F, have a variety of notable
properties. We will concentrate on SrNb$_6$O$_{16}$, which has a
high symmetry structure consisting of layers of
Nb$_6$O$_{10}$ spaced by SrO$_6$ layers.  The material has a
non-reversible in-plane polarization, but the switchable portion of
the polarization consists of out-of-plane buckling in the
Nb$_6$O$_{10}$ layers.  Because of the large number of atoms in the
unit cell, there are a variety of unstable polar and non-polar bucklings. SrNb$_6$O$_{16}$ has anomalously large effective
charges in the $z$-direction of 9 to 11 on the Nb atoms and $-7$ to $-9$
on the O atoms.  In the $xy$-plane, the effective charges are much lower, and Sr has an effective charge of 2.4. Similar anomalous
effective charges are well-known in the perovskite oxides\cite{loto}
but are not present in many alternative
ferroelectrics\cite{hyperferro,ymno3_origin}.

Despite these enormous effective charges, SrNb$_6$O$_{16}$ is actually
a candidate hyperferroelectric\cite{hyperferro}, a very rare category
of proper ferroelectrics with instabilities of both the longitudinal
optic (LO) and transverse optic (TO) modes, as shown in
table~\ref{tab:zeff}. These instabilities correspond to unstable polar
modes under both zero electric field ($\textbf{E}\!=\!0$) and zero
displacement field ($\textbf{D}\!=\!0$) boundary conditions. The polar
distortions of normal ferroelectrics are stable under
$\textbf{D}\!=\!0$ boundary conditions.

We were initially very surprised to find a hyperferroelectric with
large effective charges, as the energetic cost of producing long range
electric fields is proportional to $(Z*)^2/\epsilon$, where $\epsilon$
is the electronic dielectric constant\cite{hyperferro,hyperferro_li}.
SrNb$_6$O$_{16}$ avoids this by having three different unstable polar
modes that can mix with each other, producing modes that are still
polar, but with much smaller mode effective charges under
$\textbf{D}\!=\!0$ boundary conditions, as detailed in table
\ref{tab:zeff}. Hopefully, hyperferroelectric oxides similar to
SrNb$_6$O$_{16}$ are more common than previously thought, especially
as hyperferroelectrics have potential applications as single layer
ferroelectric devices. Further work must be done to 
understand the $\textbf{D}\!=\!0$ ground state in this material\cite{mengye}.

In our second search strategy, we considered a list about 2750
compounds from the ICSD with small unit cells and polar ground
states.  For each compound, we used the symmetry detection algorithm of
pymatgen\cite{pymatgen} and varied the tolerance factor from $10^{-6}$ to $3$,
looking for related structures with higher symmetry with otherwise
reasonable structures (\text{e.g.} no atoms on top of each other). For
compounds where new structures were found, we then relaxed both the
experimentally known polar structure and any new structures. In most
cases, the new structures were very high energy.  For example, in many
structures with covalent bonds, the new structures had broken bonds
and were unrealistic. In addition, we found a small number of compounds in
the opposite situation, where the polar phase relaxed to a non-polar structure.  These materials were either
misidentified as polar experimentally, or are not well described by the PBEsol functional. 
However, in addition to those cases, we found ten
compounds that have a low but non-zero energy difference between their polar and
non-polar structures, making them candidate ferroelectrics. We list them in
the second section of table~\ref{tab:results}, and we discuss a few interesting cases below.

First, we note that Sb$_2$WO$_6$ is a second example of a previously known
(anti-)ferroelectric/ferroelastic\cite{sbwo} that our procedure rediscovered. In
this case, the polar structure was previously characterized experimentally, and we have provided
new information about the non-polar phase.

The compound CuBiW$_2$O$_8$, as well as the five related materials
listed in the ICSD with structure type CuNb(WO$_4$)$_2\beta$, is
listed as having no symmetry in the ICSD\cite{cubiwo}. However,
according to PBEsol calculations, CuBiW$_2$O$_8$ and the related
CuYW$_2$O$_8$ relax to a higher symmetry structure with only inversion
symmetry, suggesting that these materials do not have a polar phase.
We were surprised to find so many materials possibly misidentified,
so we performed LDA\cite{lda1} calculations to verify our result. We instead found that CuBiW$_2$O$_8$ (but not CuYW$_2$O$_8$) does have polar distortion,
with a very small energy difference and polarization, as listed in
table \ref{tab:results}. This functional-dependent behavior is the
opposite of the typical behavior of LDA and GGA functionals, where GGA
tends to overestimate polar distortions, while LDA
underestimates. Further work may be necessary to reveal whether this
class of materials is really polar/ferroelectric.

\begin{figure}
\includegraphics[width=2.6in]{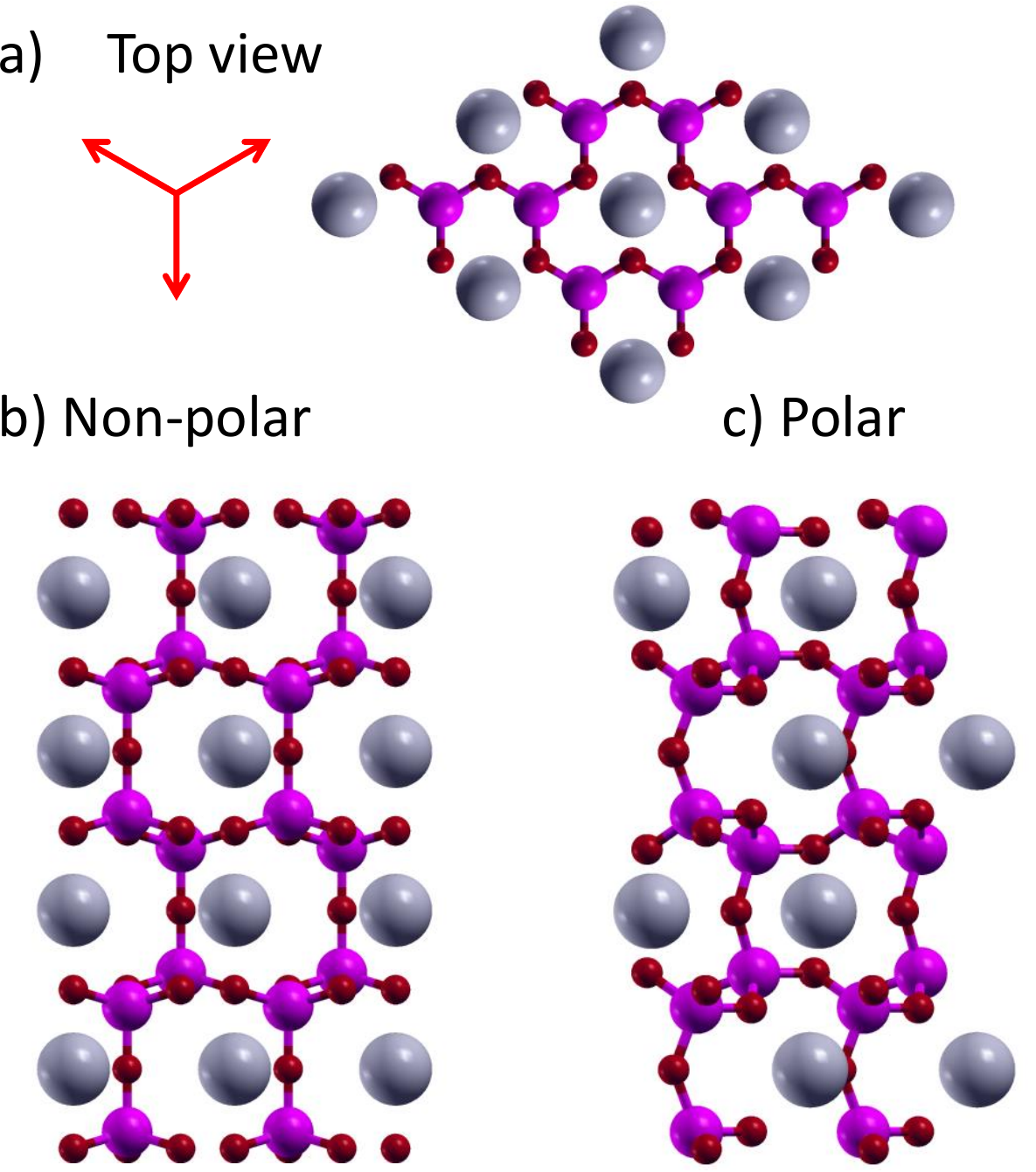}% Here is how to import EPS art
\caption{\label{fig:pbalo} Atomic positions of PbAl$_2$O$_4$. a-b) Top
  and side view of non-polar phase. c) Side view of polar phase. Red
  arrows in a) show three equivalent polarization directions.  Large
  gray atoms are Pb, medium magenta atoms are Al, and small red atoms
  are O.}
\end{figure}

The pair of compounds PbGa$_2$O$_4$ and PbAl$_2$O$_4$ consist of
(Al,Ga)O$_4$ tetrahedra separated by Pb atoms\cite{pbalo}, as shown
for PbAl$_2$O$_4$ in Fig.~\ref{fig:pbalo}. Their high symmetry
structures have a three-fold rotation axis, and the polarization,
which is due to a collective tilting of the tetrahedra, points in one
of three equivalent in-plane directions, as shown in
Fig.~\ref{fig:pbalo}a.  The structure with
180$^{\circ}$ reversed polarization does not have the same energy,
which can be understood by examining the symmetry invariant free
energy of the two-dimensional $\Gamma_5$ distortion up to fourth
order:
\begin{eqnarray}
F =  c_2 (\Gamma_{5a}^2 + \Gamma_{5b}^2) +  c_3 (\Gamma_{5a}^3 - 3 \Gamma_{5a} \Gamma_{5b}^2) \nonumber \\ 
   + c_4 (\Gamma_{5a}^4 + 2 \Gamma_{5a}^2 \Gamma_{5b}^2 + \Gamma_{5b}^4).
\end{eqnarray}
In this notation the ground state polar structure corresponds to
$\Gamma_{5a}\!>\!0$, $\Gamma_{5b}\!=\!0$, and the constants $c_2$ and
$c_3$ are negative, while $c_4$ is positive.  Unlike a typical proper
ferroelectric, there is a third-order term, which explains why the
structure with reversed polarization is inequivalent.  The reversed
polarization direction structure is only 9 meV/atom higher in energy
higher than the ground state for PbAl$_2$O$_4$, and represents the
transition state between two stable polarization directions.  Therefore, it is
possible to switch the polarization of PbGa$_2$O$_4$ and PbAl$_2$O$_4$
by rotating it 120$^{\circ}$ while avoiding
the higher energy high-symmetry structure, resulting in a barrier that is eight times lower in energy.

\begin{figure}
\includegraphics[width=3.2in]{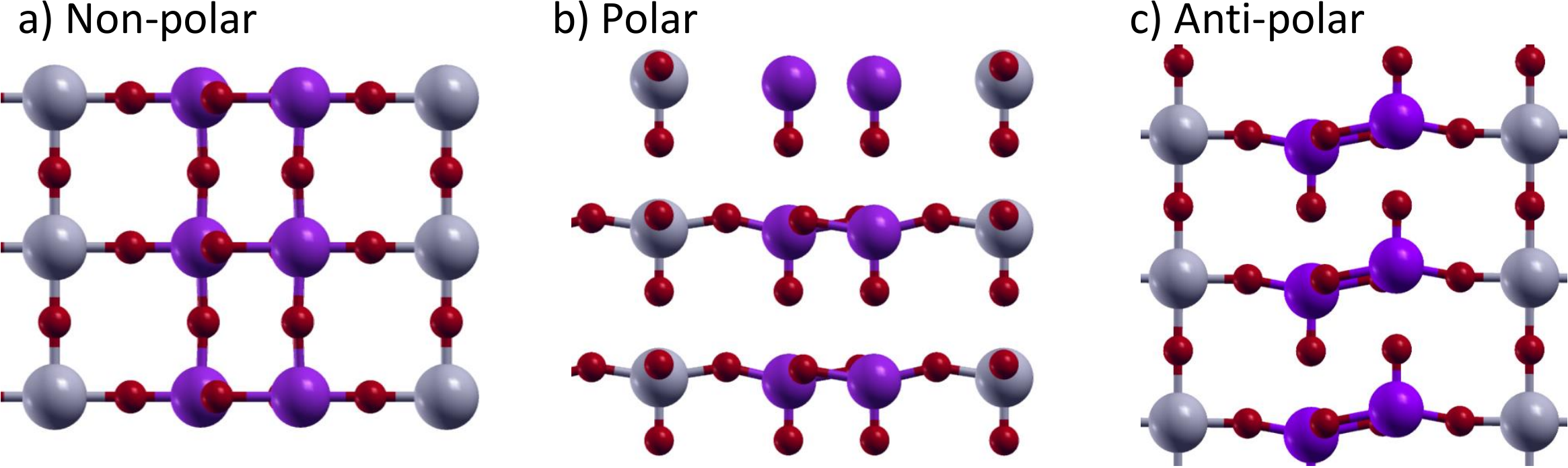}% Here is how to import EPS art
\caption{\label{fig:vmoo} a) High-symmetry, b) polar, and c) antipolar phases of V$_2$MoO$_8$. The medium gray and magenta atoms are Mo and V, respectively, and the small red atoms are O.}
\end{figure}

The layered structure of V$_2$MoO$_8$\cite{vmoo_fe} is interesting for possible
applications because it has an enormous polarization of 108 $\mu C /
cm^2$, comparable to the largest polarizations in perovskites\cite{large_polarization}.  As
shown in Fig.~\ref{fig:vmoo}, this material actually has three low
energy structures: the high-symmetry phase, the polar phase, and an
antipolar phase\cite{vmoo_af1,vmoo_af2}, which is also seen experimentally and which is 48
meV/atom lower in energy than the polar phase. Because the ground
state of V$_2$MoO$_8$ is antipolar with a competing polar phase, this
material is better characterized as a candidate antiferroelectric. The
very large polarization of the polar phase of this material could make
V$_2$MoO$_8$ useful in applications like antiferroelectric energy storage, where a
large polarization jump is desirable.

\begin{figure}
\includegraphics[width=3.0in]{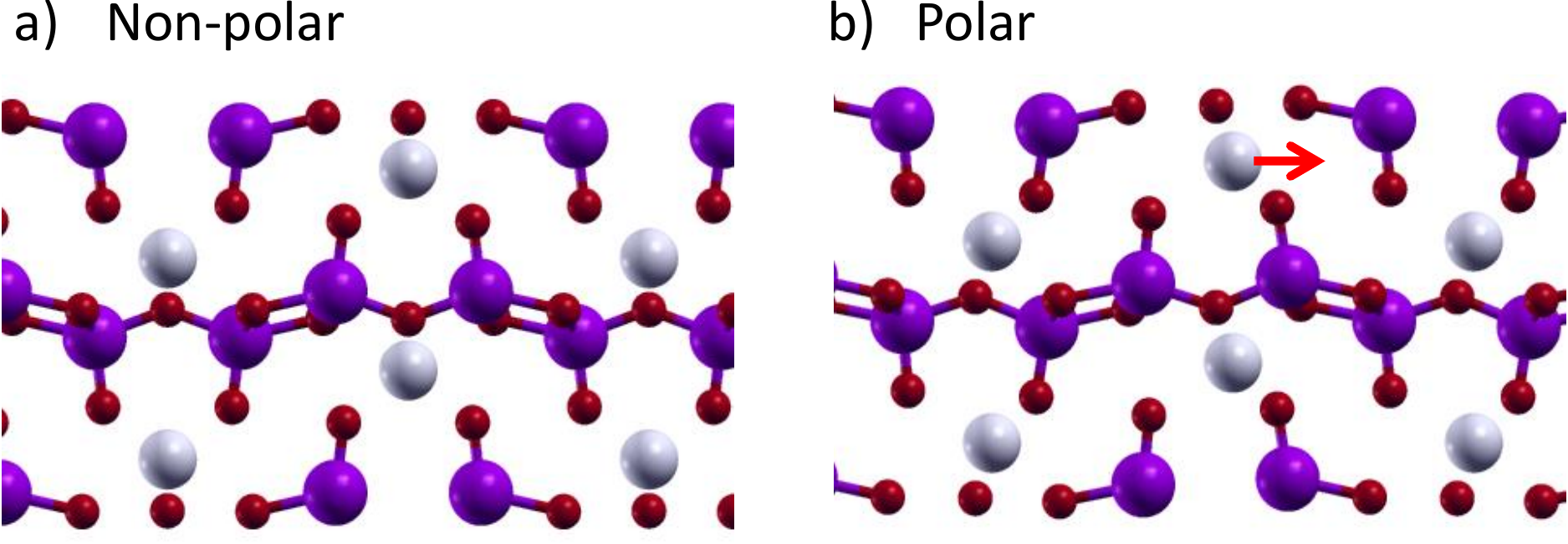}% Here is how to import EPS art
\caption{\label{fig:livo} a) Non-polar and b) polar phases of LiV$_2$O$_5$. The red arrow in b) highlights off-centering of Li. Light gray atoms are Li, magenta atoms are V, and small red atoms are O.}
\end{figure}

Finally, we note LiV$_2$O$_5$, which consists of layers of V$_2$O$_5$
spaced by Li ions, as shown in Fig.~\ref{fig:livo}, and which has
shown experimental evidence of a phase transition\cite{livo}. The
ferroelectric polarization is driven by the off-centering of Li
atoms. According to PBEsol, the material is non-magnetic;
however, we also examined various magnetic orderings using DFT+U\cite{ldaplusU,
  ldaplusU_simplified, dftU_simp}, with
a U value of 3 eV on the V $d$-states.  These calculations, which we
expect to be more accurate for a material with partially occupied 3$d$
orbitals, shows that LiV$_2$O$_5$ is a ferromagnetic insulator with a
gap of 0.37 eV in the non-polar phase and 1.00 eV in the polar
phase, making LiV$_2$O$_5$ a candidate multiferroic.

In conclusion, we have performed a high-throughput first principles
search for new ferroelectric compounds.  We use two search strategies:
1) starting with known non-polar structures and looking for polar
distortions, and 2) starting with known polar structures and looking
for related high-symmetry structures.  We discover 16 candidate
materials with a variety of interesting properties, including very
large polarizations, hyperferroelectricity, antiferroelectricity, and
multiferroism. We hope that the particular compounds found in this
work, as well as the variety of new chemistries and structures that
lead to polar distortions, will be useful in future work understanding and designing ferroelectrics.  
In addition, the high-throughout search techniques
used in this work should be applicable to future studies of structural
phase transitions.

\begin{acknowledgments}
We wish to acknowledge discussions with Igor Levin and Eric J. Cockayne, as well as help with the
ICSD from Vicky Karen and Xiang Li.
\end{acknowledgments}

%\renewcommand*{\bibfont}{\scriptsize }
%\bibliographystyle{thermo_oxides}
%\bibliography{thermo_oxides} % Produces the bibliography via BibTeX.

\begin{thebibliography}{65}
\expandafter\ifx\csname natexlab\endcsname\relax\def\natexlab#1{#1}\fi
\expandafter\ifx\csname bibnamefont\endcsname\relax
  \def\bibnamefont#1{#1}\fi
\expandafter\ifx\csname bibfnamefont\endcsname\relax
  \def\bibfnamefont#1{#1}\fi
\expandafter\ifx\csname citenamefont\endcsname\relax
  \def\citenamefont#1{#1}\fi
\expandafter\ifx\csname url\endcsname\relax
  \def\url#1{\texttt{#1}}\fi
\expandafter\ifx\csname urlprefix\endcsname\relax\def\urlprefix{URL }\fi
\providecommand{\bibinfo}[2]{#2}
\providecommand{\eprint}[2][]{\url{#2}}

\bibitem[{\citenamefont{Bennett et~al.}(2012)\citenamefont{Bennett, Garrity,
  Rabe, and Vanderbilt}}]{abcferro}
\bibinfo{author}{\bibfnamefont{J.~W.} \bibnamefont{Bennett}},
  \bibinfo{author}{\bibfnamefont{K.~F.} \bibnamefont{Garrity}},
  \bibinfo{author}{\bibfnamefont{K.~M.} \bibnamefont{Rabe}}, \bibnamefont{and}
  \bibinfo{author}{\bibfnamefont{D.}~\bibnamefont{Vanderbilt}},
  \bibinfo{journal}{Phys. Rev. Lett.} \textbf{\bibinfo{volume}{109}},
  \bibinfo{pages}{167602} (\bibinfo{year}{2012}).

\bibitem[{\citenamefont{Bennett et~al.}(2013)\citenamefont{Bennett, Garrity,
  Rabe, and Vanderbilt}}]{abcantiferro}
\bibinfo{author}{\bibfnamefont{J.~W.} \bibnamefont{Bennett}},
  \bibinfo{author}{\bibfnamefont{K.~F.} \bibnamefont{Garrity}},
  \bibinfo{author}{\bibfnamefont{K.~M.} \bibnamefont{Rabe}}, \bibnamefont{and}
  \bibinfo{author}{\bibfnamefont{D.}~\bibnamefont{Vanderbilt}},
  \bibinfo{journal}{Phys. Rev. Lett.} \textbf{\bibinfo{volume}{110}},
  \bibinfo{pages}{017603} (\bibinfo{year}{2013}).

\bibitem[{\citenamefont{Roy et~al.}(2012)\citenamefont{Roy, Bennett, Rabe, and
  Vanderbilt}}]{halfheusler}
\bibinfo{author}{\bibfnamefont{A.}~\bibnamefont{Roy}},
  \bibinfo{author}{\bibfnamefont{J.~W.} \bibnamefont{Bennett}},
  \bibinfo{author}{\bibfnamefont{K.~M.} \bibnamefont{Rabe}}, \bibnamefont{and}
  \bibinfo{author}{\bibfnamefont{D.}~\bibnamefont{Vanderbilt}},
  \bibinfo{journal}{Phys. Rev. Lett.} \textbf{\bibinfo{volume}{109}},
  \bibinfo{pages}{037602} (\bibinfo{year}{2012}).

\bibitem[{\citenamefont{Garrity
  et~al.}(2014{\natexlab{a}})\citenamefont{Garrity, Rabe, and
  Vanderbilt}}]{hyperferro}
\bibinfo{author}{\bibfnamefont{K.~F.} \bibnamefont{Garrity}},
  \bibinfo{author}{\bibfnamefont{K.~M.} \bibnamefont{Rabe}}, \bibnamefont{and}
  \bibinfo{author}{\bibfnamefont{D.}~\bibnamefont{Vanderbilt}},
  \bibinfo{journal}{Phys. Rev. Lett.} \textbf{\bibinfo{volume}{112}},
  \bibinfo{pages}{127601} (\bibinfo{year}{2014}{\natexlab{a}}).

\bibitem[{\citenamefont{Benedek and Fennie}(2011)}]{hybridimproper}
\bibinfo{author}{\bibfnamefont{N.~A.} \bibnamefont{Benedek}} \bibnamefont{and}
  \bibinfo{author}{\bibfnamefont{C.~J.} \bibnamefont{Fennie}},
  \bibinfo{journal}{Phys. Rev. Lett.} \textbf{\bibinfo{volume}{106}},
  \bibinfo{pages}{107204} (\bibinfo{year}{2011}).

\bibitem[{\citenamefont{Van~Aken et~al.}(2004)\citenamefont{Van~Aken, Palstra,
  Filippetti, and Spaldin}}]{ymno3_origin}
\bibinfo{author}{\bibfnamefont{B.~B.} \bibnamefont{Van~Aken}},
  \bibinfo{author}{\bibfnamefont{T.~T.} \bibnamefont{Palstra}},
  \bibinfo{author}{\bibfnamefont{A.}~\bibnamefont{Filippetti}},
  \bibnamefont{and} \bibinfo{author}{\bibfnamefont{N.~A.}
  \bibnamefont{Spaldin}}, \bibinfo{journal}{Nature materials}
  \textbf{\bibinfo{volume}{3}}, \bibinfo{pages}{164} (\bibinfo{year}{2004}).

\bibitem[{\citenamefont{Fennie and Rabe}(2005)}]{ymno3}
\bibinfo{author}{\bibfnamefont{C.~J.} \bibnamefont{Fennie}} \bibnamefont{and}
  \bibinfo{author}{\bibfnamefont{K.~M.} \bibnamefont{Rabe}},
  \bibinfo{journal}{Physical Review B} \textbf{\bibinfo{volume}{72}},
  \bibinfo{pages}{100103} (\bibinfo{year}{2005}).

\bibitem[{\citenamefont{Benedek}(2014)}]{dionjacob}
\bibinfo{author}{\bibfnamefont{N.~A.} \bibnamefont{Benedek}},
  \bibinfo{journal}{Inorganic Chemistry} \textbf{\bibinfo{volume}{53}},
  \bibinfo{pages}{3769} (\bibinfo{year}{2014}), \bibinfo{note}{pMID: 24678981}.

\bibitem[{\citenamefont{Reyes-Lillo et~al.}(2014)\citenamefont{Reyes-Lillo,
  Garrity, and Rabe}}]{zro2}
\bibinfo{author}{\bibfnamefont{S.~E.} \bibnamefont{Reyes-Lillo}},
  \bibinfo{author}{\bibfnamefont{K.~F.} \bibnamefont{Garrity}},
  \bibnamefont{and} \bibinfo{author}{\bibfnamefont{K.~M.} \bibnamefont{Rabe}},
  \bibinfo{journal}{Phys. Rev. B} \textbf{\bibinfo{volume}{90}},
  \bibinfo{pages}{140103} (\bibinfo{year}{2014}),
  \urlprefix\url{http://link.aps.org/doi/10.1103/PhysRevB.90.140103}.

\bibitem[{\citenamefont{Fennie and Rabe}(2006)}]{eutio3}
\bibinfo{author}{\bibfnamefont{C.~J.} \bibnamefont{Fennie}} \bibnamefont{and}
  \bibinfo{author}{\bibfnamefont{K.~M.} \bibnamefont{Rabe}},
  \bibinfo{journal}{Phys. Rev. Lett.} \textbf{\bibinfo{volume}{97}},
  \bibinfo{pages}{267602} (\bibinfo{year}{2006}),
  \urlprefix\url{http://link.aps.org/doi/10.1103/PhysRevLett.97.267602}.

\bibitem[{\citenamefont{Oh et~al.}(2015)\citenamefont{Oh, Luo, Huang, Wang, and
  Cheong}}]{switch_hybrid_improper}
\bibinfo{author}{\bibfnamefont{Y.~S.} \bibnamefont{Oh}},
  \bibinfo{author}{\bibfnamefont{X.}~\bibnamefont{Luo}},
  \bibinfo{author}{\bibfnamefont{F.-T.} \bibnamefont{Huang}},
  \bibinfo{author}{\bibfnamefont{Y.}~\bibnamefont{Wang}}, \bibnamefont{and}
  \bibinfo{author}{\bibfnamefont{S.-W.} \bibnamefont{Cheong}},
  \bibinfo{journal}{Nature materials} \textbf{\bibinfo{volume}{14}},
  \bibinfo{pages}{407} (\bibinfo{year}{2015}).

\bibitem[{\citenamefont{Lee et~al.}(2010)\citenamefont{Lee, Fang, Vlahos, Ke,
  Jung, Kourkoutis, Kim, Ryan, Heeg, Roeckerath et~al.}}]{eutio3_expt}
\bibinfo{author}{\bibfnamefont{J.~H.} \bibnamefont{Lee}},
  \bibinfo{author}{\bibfnamefont{L.}~\bibnamefont{Fang}},
  \bibinfo{author}{\bibfnamefont{E.}~\bibnamefont{Vlahos}},
  \bibinfo{author}{\bibfnamefont{X.}~\bibnamefont{Ke}},
  \bibinfo{author}{\bibfnamefont{Y.~W.} \bibnamefont{Jung}},
  \bibinfo{author}{\bibfnamefont{L.~F.} \bibnamefont{Kourkoutis}},
  \bibinfo{author}{\bibfnamefont{J.-W.} \bibnamefont{Kim}},
  \bibinfo{author}{\bibfnamefont{P.~J.} \bibnamefont{Ryan}},
  \bibinfo{author}{\bibfnamefont{T.}~\bibnamefont{Heeg}},
  \bibinfo{author}{\bibfnamefont{M.}~\bibnamefont{Roeckerath}},
  \bibnamefont{et~al.}, \bibinfo{journal}{Nature}
  \textbf{\bibinfo{volume}{466}}, \bibinfo{pages}{954} (\bibinfo{year}{2010}).

\bibitem[{\citenamefont{Choi et~al.}(2010)\citenamefont{Choi, Horibe, Yi, Choi,
  Wu, and Cheong}}]{ymno3_domainwalls}
\bibinfo{author}{\bibfnamefont{T.}~\bibnamefont{Choi}},
  \bibinfo{author}{\bibfnamefont{Y.}~\bibnamefont{Horibe}},
  \bibinfo{author}{\bibfnamefont{H.}~\bibnamefont{Yi}},
  \bibinfo{author}{\bibfnamefont{Y.}~\bibnamefont{Choi}},
  \bibinfo{author}{\bibfnamefont{W.}~\bibnamefont{Wu}}, \bibnamefont{and}
  \bibinfo{author}{\bibfnamefont{S.-W.} \bibnamefont{Cheong}},
  \bibinfo{journal}{Nature materials} \textbf{\bibinfo{volume}{9}},
  \bibinfo{pages}{253} (\bibinfo{year}{2010}).

\bibitem[{\citenamefont{Lee and Rabe}(2010)}]{srmno3}
\bibinfo{author}{\bibfnamefont{J.~H.} \bibnamefont{Lee}} \bibnamefont{and}
  \bibinfo{author}{\bibfnamefont{K.~M.} \bibnamefont{Rabe}},
  \bibinfo{journal}{Phys. Rev. Lett.} \textbf{\bibinfo{volume}{104}},
  \bibinfo{pages}{207204} (\bibinfo{year}{2010}),
  \urlprefix\url{http://link.aps.org/doi/10.1103/PhysRevLett.104.207204}.

\bibitem[{\citenamefont{Garrity et~al.}(2013)\citenamefont{Garrity, Kakekhani,
  Kolpak, and Ismail-Beigi}}]{garrity_fe_surface}
\bibinfo{author}{\bibfnamefont{K.}~\bibnamefont{Garrity}},
  \bibinfo{author}{\bibfnamefont{A.}~\bibnamefont{Kakekhani}},
  \bibinfo{author}{\bibfnamefont{A.}~\bibnamefont{Kolpak}}, \bibnamefont{and}
  \bibinfo{author}{\bibfnamefont{S.}~\bibnamefont{Ismail-Beigi}},
  \bibinfo{journal}{Physical Review B} \textbf{\bibinfo{volume}{88}},
  \bibinfo{pages}{045401} (\bibinfo{year}{2013}).

\bibitem[{\citenamefont{Garrity et~al.}(2010)\citenamefont{Garrity, Kolpak,
  Ismail-Beigi, and Altman}}]{ferro_surf_review}
\bibinfo{author}{\bibfnamefont{K.}~\bibnamefont{Garrity}},
  \bibinfo{author}{\bibfnamefont{A.~M.} \bibnamefont{Kolpak}},
  \bibinfo{author}{\bibfnamefont{S.}~\bibnamefont{Ismail-Beigi}},
  \bibnamefont{and} \bibinfo{author}{\bibfnamefont{E.~I.}
  \bibnamefont{Altman}}, \bibinfo{journal}{Advanced Materials}
  \textbf{\bibinfo{volume}{22}}, \bibinfo{pages}{2969} (\bibinfo{year}{2010}),
  ISSN \bibinfo{issn}{1521-4095},
  \urlprefix\url{http://dx.doi.org/10.1002/adma.200903723}.

\bibitem[{\citenamefont{Young and Rappe}(2012)}]{shiftcurrent}
\bibinfo{author}{\bibfnamefont{S.~M.} \bibnamefont{Young}} \bibnamefont{and}
  \bibinfo{author}{\bibfnamefont{A.~M.} \bibnamefont{Rappe}},
  \bibinfo{journal}{Physical review letters} \textbf{\bibinfo{volume}{109}},
  \bibinfo{pages}{116601} (\bibinfo{year}{2012}).

\bibitem[{\citenamefont{Reiner et~al.}(2010)\citenamefont{Reiner, Kolpak,
  Segal, Garrity, Ismail-Beigi, Ahn, and Walker}}]{silicon_oxide}
\bibinfo{author}{\bibfnamefont{J.~W.} \bibnamefont{Reiner}},
  \bibinfo{author}{\bibfnamefont{A.~M.} \bibnamefont{Kolpak}},
  \bibinfo{author}{\bibfnamefont{Y.}~\bibnamefont{Segal}},
  \bibinfo{author}{\bibfnamefont{K.~F.} \bibnamefont{Garrity}},
  \bibinfo{author}{\bibfnamefont{S.}~\bibnamefont{Ismail-Beigi}},
  \bibinfo{author}{\bibfnamefont{C.~H.} \bibnamefont{Ahn}}, \bibnamefont{and}
  \bibinfo{author}{\bibfnamefont{F.~J.} \bibnamefont{Walker}},
  \bibinfo{journal}{Advanced Materials} \textbf{\bibinfo{volume}{22}},
  \bibinfo{pages}{2919} (\bibinfo{year}{2010}), ISSN \bibinfo{issn}{1521-4095},
  \urlprefix\url{http://dx.doi.org/10.1002/adma.200904306}.

\bibitem[{\citenamefont{Müller et~al.}(2015)\citenamefont{Müller, Polakowski,
  Mueller, and Mikolajick}}]{hfo2_devices}
\bibinfo{author}{\bibfnamefont{J.}~\bibnamefont{Müller}},
  \bibinfo{author}{\bibfnamefont{P.}~\bibnamefont{Polakowski}},
  \bibinfo{author}{\bibfnamefont{S.}~\bibnamefont{Mueller}}, \bibnamefont{and}
  \bibinfo{author}{\bibfnamefont{T.}~\bibnamefont{Mikolajick}},
  \bibinfo{journal}{ECS Journal of Solid State Science and Technology}
  \textbf{\bibinfo{volume}{4}}, \bibinfo{pages}{N30} (\bibinfo{year}{2015}),
  \eprint{http://jss.ecsdl.org/content/4/5/N30.full.pdf+html},
  \urlprefix\url{http://jss.ecsdl.org/content/4/5/N30.abstract}.

\bibitem[{\citenamefont{Hill}(2000)}]{few_multiferroics}
\bibinfo{author}{\bibfnamefont{N.~A.} \bibnamefont{Hill}},
  \bibinfo{journal}{The Journal of Physical Chemistry B}
  \textbf{\bibinfo{volume}{104}}, \bibinfo{pages}{6694} (\bibinfo{year}{2000}).

\bibitem[{\citenamefont{Jain et~al.}(2011)\citenamefont{Jain, Hautier, Moore,
  Ong, Fischer, Mueller, Persson, and Ceder}}]{compmatsci}
\bibinfo{author}{\bibfnamefont{A.}~\bibnamefont{Jain}},
  \bibinfo{author}{\bibfnamefont{G.}~\bibnamefont{Hautier}},
  \bibinfo{author}{\bibfnamefont{C.~J.} \bibnamefont{Moore}},
  \bibinfo{author}{\bibfnamefont{S.~P.} \bibnamefont{Ong}},
  \bibinfo{author}{\bibfnamefont{C.~C.} \bibnamefont{Fischer}},
  \bibinfo{author}{\bibfnamefont{T.}~\bibnamefont{Mueller}},
  \bibinfo{author}{\bibfnamefont{K.~A.} \bibnamefont{Persson}},
  \bibnamefont{and} \bibinfo{author}{\bibfnamefont{G.}~\bibnamefont{Ceder}},
  \bibinfo{journal}{Comput. Mater. Sci.} \textbf{\bibinfo{volume}{50}},
  \bibinfo{pages}{2295 } (\bibinfo{year}{2011}), ISSN
  \bibinfo{issn}{0927-0256}.

\bibitem[{\citenamefont{Morgan et~al.}(2005)\citenamefont{Morgan, Ceder, and
  Curtarolo}}]{datamine}
\bibinfo{author}{\bibfnamefont{D.}~\bibnamefont{Morgan}},
  \bibinfo{author}{\bibfnamefont{G.}~\bibnamefont{Ceder}}, \bibnamefont{and}
  \bibinfo{author}{\bibfnamefont{S.}~\bibnamefont{Curtarolo}},
  \bibinfo{journal}{Meas. Sci. \& Tech.} \textbf{\bibinfo{volume}{16}},
  \bibinfo{pages}{296} (\bibinfo{year}{2005}), ISSN \bibinfo{issn}{0957-0233}.

\bibitem[{\citenamefont{R.~Akbarzadeh et~al.}(2007)\citenamefont{R.~Akbarzadeh,
  Ozoliņš, and Wolverton}}]{wolverton}
\bibinfo{author}{\bibfnamefont{A.}~\bibnamefont{R.~Akbarzadeh}},
  \bibinfo{author}{\bibfnamefont{V.}~\bibnamefont{Ozoliņš}},
  \bibnamefont{and}
  \bibinfo{author}{\bibfnamefont{C.}~\bibnamefont{Wolverton}},
  \bibinfo{journal}{Adv. Mater.} \textbf{\bibinfo{volume}{19}},
  \bibinfo{pages}{3233} (\bibinfo{year}{2007}), ISSN \bibinfo{issn}{1521-4095}.

\bibitem[{\citenamefont{Jain et~al.}(2013)\citenamefont{Jain, Ong, Hautier,
  Chen, Richards, Dacek, Cholia, Gunter, Skinner, Ceder
  et~al.}}]{materialsproject}
\bibinfo{author}{\bibfnamefont{A.}~\bibnamefont{Jain}},
  \bibinfo{author}{\bibfnamefont{S.~P.} \bibnamefont{Ong}},
  \bibinfo{author}{\bibfnamefont{G.}~\bibnamefont{Hautier}},
  \bibinfo{author}{\bibfnamefont{W.}~\bibnamefont{Chen}},
  \bibinfo{author}{\bibfnamefont{W.~D.} \bibnamefont{Richards}},
  \bibinfo{author}{\bibfnamefont{S.}~\bibnamefont{Dacek}},
  \bibinfo{author}{\bibfnamefont{S.}~\bibnamefont{Cholia}},
  \bibinfo{author}{\bibfnamefont{D.}~\bibnamefont{Gunter}},
  \bibinfo{author}{\bibfnamefont{D.}~\bibnamefont{Skinner}},
  \bibinfo{author}{\bibfnamefont{G.}~\bibnamefont{Ceder}},
  \bibnamefont{et~al.}, \bibinfo{journal}{APL Materials}
  \textbf{\bibinfo{volume}{1}}, \bibinfo{pages}{011002} (\bibinfo{year}{2013}),
  ISSN \bibinfo{issn}{2166532X},
  \urlprefix\url{http://link.aip.org/link/AMPADS/v1/i1/p011002/s1\&Agg=doi}.

\bibitem[{\citenamefont{Curtarolo et~al.}(2012)\citenamefont{Curtarolo,
  Setyawan, Wang, Xue, Yang, Taylor, Nelson, Hart, Sanvito, Buongiorno-Nardelli
  et~al.}}]{aflowlib}
\bibinfo{author}{\bibfnamefont{S.}~\bibnamefont{Curtarolo}},
  \bibinfo{author}{\bibfnamefont{W.}~\bibnamefont{Setyawan}},
  \bibinfo{author}{\bibfnamefont{S.}~\bibnamefont{Wang}},
  \bibinfo{author}{\bibfnamefont{J.}~\bibnamefont{Xue}},
  \bibinfo{author}{\bibfnamefont{K.}~\bibnamefont{Yang}},
  \bibinfo{author}{\bibfnamefont{R.~H.} \bibnamefont{Taylor}},
  \bibinfo{author}{\bibfnamefont{L.~J.} \bibnamefont{Nelson}},
  \bibinfo{author}{\bibfnamefont{G.~L.} \bibnamefont{Hart}},
  \bibinfo{author}{\bibfnamefont{S.}~\bibnamefont{Sanvito}},
  \bibinfo{author}{\bibfnamefont{M.}~\bibnamefont{Buongiorno-Nardelli}},
  \bibnamefont{et~al.}, \bibinfo{journal}{Computational Materials Science}
  \textbf{\bibinfo{volume}{58}}, \bibinfo{pages}{227 } (\bibinfo{year}{2012}),
  ISSN \bibinfo{issn}{0927-0256},
  \urlprefix\url{http://www.sciencedirect.com/science/article/pii/S0927025612000687}.

\bibitem[{\citenamefont{Saal et~al.}(2013)\citenamefont{Saal, Kirklin, Aykol,
  Meredig, and Wolverton}}]{oqmd}
\bibinfo{author}{\bibfnamefont{J.~E.} \bibnamefont{Saal}},
  \bibinfo{author}{\bibfnamefont{S.}~\bibnamefont{Kirklin}},
  \bibinfo{author}{\bibfnamefont{M.}~\bibnamefont{Aykol}},
  \bibinfo{author}{\bibfnamefont{B.}~\bibnamefont{Meredig}}, \bibnamefont{and}
  \bibinfo{author}{\bibfnamefont{C.}~\bibnamefont{Wolverton}},
  \bibinfo{journal}{Jom} \textbf{\bibinfo{volume}{65}}, \bibinfo{pages}{1501}
  (\bibinfo{year}{2013}).

\bibitem[{\citenamefont{Bennett and Rabe}(2012)}]{joedatabase}
\bibinfo{author}{\bibfnamefont{J.~W.} \bibnamefont{Bennett}} \bibnamefont{and}
  \bibinfo{author}{\bibfnamefont{K.~M.} \bibnamefont{Rabe}},
  \bibinfo{journal}{J. Solid State Chem.} \textbf{\bibinfo{volume}{195}},
  \bibinfo{pages}{21} (\bibinfo{year}{2012}).

\bibitem[{\citenamefont{Bennett}(2012)}]{joe_icsd}
\bibinfo{author}{\bibfnamefont{J.~W.} \bibnamefont{Bennett}},
  \bibinfo{journal}{Physics Procedia} \textbf{\bibinfo{volume}{34}},
  \bibinfo{pages}{14 } (\bibinfo{year}{2012}), ISSN \bibinfo{issn}{1875-3892},
  \bibinfo{note}{proceedings of the 25th Workshop on Computer Simulation
  Studies in Condensed Matter Physics},
  \urlprefix\url{http://www.sciencedirect.com/science/article/pii/S1875389212013168}.

\bibitem[{\citenamefont{Abrahams}(1988)}]{abrahams1}
\bibinfo{author}{\bibfnamefont{S.~C.} \bibnamefont{Abrahams}},
  \bibinfo{journal}{Acta Crystallographica Section B}
  \textbf{\bibinfo{volume}{44}}, \bibinfo{pages}{585} (\bibinfo{year}{1988}),
  \urlprefix\url{http://dx.doi.org/10.1107/S0108768188010110}.

\bibitem[{\citenamefont{Abrahams}(1996)}]{abrahams2}
\bibinfo{author}{\bibfnamefont{S.~C.} \bibnamefont{Abrahams}},
  \bibinfo{journal}{Acta Crystallographica Section B}
  \textbf{\bibinfo{volume}{52}}, \bibinfo{pages}{790} (\bibinfo{year}{1996}),
  \urlprefix\url{http://dx.doi.org/10.1107/S0108768196004594}.

\bibitem[{\citenamefont{Abrahams}(2006)}]{abrahams3}
\bibinfo{author}{\bibfnamefont{S.~C.} \bibnamefont{Abrahams}},
  \bibinfo{journal}{Acta Crystallographica Section B}
  \textbf{\bibinfo{volume}{62}}, \bibinfo{pages}{26} (\bibinfo{year}{2006}),
  \urlprefix\url{http://dx.doi.org/10.1107/S0108768105040577}.

\bibitem[{\citenamefont{Atuchin et~al.}(2004)\citenamefont{Atuchin, Kidyarov,
  and Pervukhina}}]{acentric}
\bibinfo{author}{\bibfnamefont{V.}~\bibnamefont{Atuchin}},
  \bibinfo{author}{\bibfnamefont{B.}~\bibnamefont{Kidyarov}}, \bibnamefont{and}
  \bibinfo{author}{\bibfnamefont{N.}~\bibnamefont{Pervukhina}},
  \bibinfo{journal}{Computational Materials Science}
  \textbf{\bibinfo{volume}{30}}, \bibinfo{pages}{411 } (\bibinfo{year}{2004}),
  ISSN \bibinfo{issn}{0927-0256}, \bibinfo{note}{theory, modeling and
  simulation of materials for advanced technologies: Proceedings of the
  International Conference on Materials for Advanced Technologies (ICMAT 2003)
  and \{IUMRS\} International Conference in Asia (IUMRS-ICA 2003)},
  \urlprefix\url{http://www.sciencedirect.com/science/article/pii/S0927025604001272}.

\bibitem[{\citenamefont{Halasyamani†
  et~al.}(1998)\citenamefont{Halasyamani†, , and
  Poeppelmeier*}}]{noncentro_oxides}
\bibinfo{author}{\bibfnamefont{P.~S.} \bibnamefont{Halasyamani†}}, ,
  \bibnamefont{and} \bibinfo{author}{\bibfnamefont{K.~R.}
  \bibnamefont{Poeppelmeier*}}, \bibinfo{journal}{Chemistry of Materials}
  \textbf{\bibinfo{volume}{10}}, \bibinfo{pages}{2753} (\bibinfo{year}{1998}),
  \eprint{http://dx.doi.org/10.1021/cm980140w},
  \urlprefix\url{http://dx.doi.org/10.1021/cm980140w}.

\bibitem[{\citenamefont{Bennett et~al.}(2011)\citenamefont{Bennett, Grinberg,
  Davies, and Rappe}}]{pb_free}
\bibinfo{author}{\bibfnamefont{J.~W.} \bibnamefont{Bennett}},
  \bibinfo{author}{\bibfnamefont{I.}~\bibnamefont{Grinberg}},
  \bibinfo{author}{\bibfnamefont{P.~K.} \bibnamefont{Davies}},
  \bibnamefont{and} \bibinfo{author}{\bibfnamefont{A.~M.} \bibnamefont{Rappe}},
  \bibinfo{journal}{Phys. Rev. B} \textbf{\bibinfo{volume}{83}},
  \bibinfo{pages}{144112} (\bibinfo{year}{2011}),
  \urlprefix\url{http://link.aps.org/doi/10.1103/PhysRevB.83.144112}.

\bibitem[{\citenamefont{Garrity}(2016)}]{mythermoelectrics}
\bibinfo{author}{\bibfnamefont{K.~F.} \bibnamefont{Garrity}},
  \bibinfo{journal}{Phys. Rev. B} \textbf{\bibinfo{volume}{94}},
  \bibinfo{pages}{045122} (\bibinfo{year}{2016}),
  \urlprefix\url{http://link.aps.org/doi/10.1103/PhysRevB.94.045122}.

\bibitem[{\citenamefont{Ong et~al.}(2013)\citenamefont{Ong, Richards, Jain,
  Hautier, Kocher, Cholia, Gunter, Chevrier, Persson, and Ceder}}]{pymatgen}
\bibinfo{author}{\bibfnamefont{S.~P.} \bibnamefont{Ong}},
  \bibinfo{author}{\bibfnamefont{W.~D.} \bibnamefont{Richards}},
  \bibinfo{author}{\bibfnamefont{A.}~\bibnamefont{Jain}},
  \bibinfo{author}{\bibfnamefont{G.}~\bibnamefont{Hautier}},
  \bibinfo{author}{\bibfnamefont{M.}~\bibnamefont{Kocher}},
  \bibinfo{author}{\bibfnamefont{S.}~\bibnamefont{Cholia}},
  \bibinfo{author}{\bibfnamefont{D.}~\bibnamefont{Gunter}},
  \bibinfo{author}{\bibfnamefont{V.~L.} \bibnamefont{Chevrier}},
  \bibinfo{author}{\bibfnamefont{K.~A.} \bibnamefont{Persson}},
  \bibnamefont{and} \bibinfo{author}{\bibfnamefont{G.}~\bibnamefont{Ceder}},
  \bibinfo{journal}{Computational Materials Science}
  \textbf{\bibinfo{volume}{68}}, \bibinfo{pages}{314 } (\bibinfo{year}{2013}),
  ISSN \bibinfo{issn}{0927-0256},
  \urlprefix\url{http://www.sciencedirect.com/science/article/pii/S0927025612006295}.

\bibitem[{\citenamefont{Hohenberg and Kohn}(1964)}]{hk}
\bibinfo{author}{\bibfnamefont{P.}~\bibnamefont{Hohenberg}} \bibnamefont{and}
  \bibinfo{author}{\bibfnamefont{W.}~\bibnamefont{Kohn}},
  \bibinfo{journal}{Phys.\ Rev.} \textbf{\bibinfo{volume}{136}},
  \bibinfo{pages}{B864} (\bibinfo{year}{1964}).

\bibitem[{\citenamefont{Kohn and Sham}(1965)}]{ks}
\bibinfo{author}{\bibfnamefont{W.}~\bibnamefont{Kohn}} \bibnamefont{and}
  \bibinfo{author}{\bibfnamefont{L.}~\bibnamefont{Sham}},
  \bibinfo{journal}{Phys.\ Rev.} \textbf{\bibinfo{volume}{140}},
  \bibinfo{pages}{A1133} (\bibinfo{year}{1965}).

\bibitem[{\citenamefont{Giannozzi and et~al.}(2009)}]{QE}
\bibinfo{author}{\bibfnamefont{P.}~\bibnamefont{Giannozzi}} \bibnamefont{and}
  \bibinfo{author}{\bibnamefont{et~al.}}, \bibinfo{journal}{J. Phys.:Condens.
  Matter} \textbf{\bibinfo{volume}{21}}, \bibinfo{pages}{395502}
  (\bibinfo{year}{2009}).

\bibitem[{\citenamefont{Vanderbilt}(1990)}]{ultrasoft}
\bibinfo{author}{\bibfnamefont{D.}~\bibnamefont{Vanderbilt}},
  \bibinfo{journal}{Phys. Rev. B} \textbf{\bibinfo{volume}{41}},
  \bibinfo{pages}{7892} (\bibinfo{year}{1990}).

\bibitem[{\citenamefont{Garrity
  et~al.}(2014{\natexlab{b}})\citenamefont{Garrity, Bennett, Rabe, and
  Vanderbilt}}]{gbrv}
\bibinfo{author}{\bibfnamefont{K.~F.} \bibnamefont{Garrity}},
  \bibinfo{author}{\bibfnamefont{J.~W.} \bibnamefont{Bennett}},
  \bibinfo{author}{\bibfnamefont{K.~M.} \bibnamefont{Rabe}}, \bibnamefont{and}
  \bibinfo{author}{\bibfnamefont{D.}~\bibnamefont{Vanderbilt}},
  \bibinfo{journal}{Comput. Mater. Sci} \textbf{\bibinfo{volume}{81}},
  \bibinfo{pages}{446} (\bibinfo{year}{2014}{\natexlab{b}}).

\bibitem[{\citenamefont{Garrity}(2015)}]{phonon_convergence}
\bibinfo{author}{\bibfnamefont{K.~F.} \bibnamefont{Garrity}},
  \emph{\bibinfo{title}{Gbrv phonon update and jth/pslibrary testing}}
  (\bibinfo{year}{2015}),
  \urlprefix\url{http://www.physics.rutgers.edu/gbrv/gbrv_phonon_update2.pdf}.

\bibitem[{\citenamefont{Perdew et~al.}(2008)\citenamefont{Perdew, Ruzsinszky,
  Csonka, Vydrov, Scuseria, Constantin, Zhou, and Burke}}]{pbesol}
\bibinfo{author}{\bibfnamefont{J.~P.} \bibnamefont{Perdew}},
  \bibinfo{author}{\bibfnamefont{A.}~\bibnamefont{Ruzsinszky}},
  \bibinfo{author}{\bibfnamefont{G.~I.} \bibnamefont{Csonka}},
  \bibinfo{author}{\bibfnamefont{O.~A.} \bibnamefont{Vydrov}},
  \bibinfo{author}{\bibfnamefont{G.~E.} \bibnamefont{Scuseria}},
  \bibinfo{author}{\bibfnamefont{L.~A.} \bibnamefont{Constantin}},
  \bibinfo{author}{\bibfnamefont{X.}~\bibnamefont{Zhou}}, \bibnamefont{and}
  \bibinfo{author}{\bibfnamefont{K.}~\bibnamefont{Burke}},
  \bibinfo{journal}{Phys. Rev. Lett.} \textbf{\bibinfo{volume}{100}},
  \bibinfo{pages}{136406} (\bibinfo{year}{2008}),
  \urlprefix\url{http://link.aps.org/doi/10.1103/PhysRevLett.100.136406}.

\bibitem[{\citenamefont{Baroni et~al.}(2001)\citenamefont{Baroni, de~Gironcoli,
  Dal~Corso, and Giannozzi}}]{dft-pt}
\bibinfo{author}{\bibfnamefont{S.}~\bibnamefont{Baroni}},
  \bibinfo{author}{\bibfnamefont{S.}~\bibnamefont{de~Gironcoli}},
  \bibinfo{author}{\bibfnamefont{A.}~\bibnamefont{Dal~Corso}},
  \bibnamefont{and}
  \bibinfo{author}{\bibfnamefont{P.}~\bibnamefont{Giannozzi}},
  \bibinfo{journal}{Rev. Mod. Phys.} \textbf{\bibinfo{volume}{73}},
  \bibinfo{pages}{515} (\bibinfo{year}{2001}),
  \urlprefix\url{http://link.aps.org/doi/10.1103/RevModPhys.73.515}.

\bibitem[{\citenamefont{King-Smith and Vanderbilt}(1993)}]{modern_polarization}
\bibinfo{author}{\bibfnamefont{R.}~\bibnamefont{King-Smith}} \bibnamefont{and}
  \bibinfo{author}{\bibfnamefont{D.}~\bibnamefont{Vanderbilt}},
  \bibinfo{journal}{Physical Review B} \textbf{\bibinfo{volume}{47}},
  \bibinfo{pages}{1651} (\bibinfo{year}{1993}).

\bibitem[{\citenamefont{Di\'eguez et~al.}(2011)\citenamefont{Di\'eguez,
  Gonz\'alez-V\'azquez, Wojde\l{}, and \'I\~niguez}}]{bifeo3}
\bibinfo{author}{\bibfnamefont{O.}~\bibnamefont{Di\'eguez}},
  \bibinfo{author}{\bibfnamefont{O.~E.} \bibnamefont{Gonz\'alez-V\'azquez}},
  \bibinfo{author}{\bibfnamefont{J.~C.} \bibnamefont{Wojde\l{}}},
  \bibnamefont{and}
  \bibinfo{author}{\bibfnamefont{J.}~\bibnamefont{\'I\~niguez}},
  \bibinfo{journal}{Phys. Rev. B} \textbf{\bibinfo{volume}{83}},
  \bibinfo{pages}{094105} (\bibinfo{year}{2011}),
  \urlprefix\url{http://link.aps.org/doi/10.1103/PhysRevB.83.094105}.

\bibitem[{\citenamefont{Zhou and
  Rabe}(2014)}]{interface_structure_determination}
\bibinfo{author}{\bibfnamefont{Y.}~\bibnamefont{Zhou}} \bibnamefont{and}
  \bibinfo{author}{\bibfnamefont{K.~M.} \bibnamefont{Rabe}},
  \bibinfo{journal}{Physical Review B} \textbf{\bibinfo{volume}{89}},
  \bibinfo{pages}{214108} (\bibinfo{year}{2014}).

\bibitem[{\citenamefont{Rabe}()}]{rabeantiferroelectricity}
\bibinfo{author}{\bibfnamefont{K.~M.} \bibnamefont{Rabe}},
  \bibinfo{journal}{Functional Metal Oxides: New Science and Novel
  Applications} pp. \bibinfo{pages}{221--244} (????).

\bibitem[{\citenamefont{Shimakawa et~al.}(2000)\citenamefont{Shimakawa, Kubo,
  Nakagawa, Goto, Kamiyama, Asano, and Izumi}}]{babitao}
\bibinfo{author}{\bibfnamefont{Y.}~\bibnamefont{Shimakawa}},
  \bibinfo{author}{\bibfnamefont{Y.}~\bibnamefont{Kubo}},
  \bibinfo{author}{\bibfnamefont{Y.}~\bibnamefont{Nakagawa}},
  \bibinfo{author}{\bibfnamefont{S.}~\bibnamefont{Goto}},
  \bibinfo{author}{\bibfnamefont{T.}~\bibnamefont{Kamiyama}},
  \bibinfo{author}{\bibfnamefont{H.}~\bibnamefont{Asano}}, \bibnamefont{and}
  \bibinfo{author}{\bibfnamefont{F.}~\bibnamefont{Izumi}},
  \bibinfo{journal}{Phys. Rev. B} \textbf{\bibinfo{volume}{61}},
  \bibinfo{pages}{6559} (\bibinfo{year}{2000}),
  \urlprefix\url{http://link.aps.org/doi/10.1103/PhysRevB.61.6559}.

\bibitem[{\citenamefont{Marinder et~al.}(1986)\citenamefont{Marinder, Wang, and
  Werner}}]{srnbo}
\bibinfo{author}{\bibfnamefont{B.}~\bibnamefont{Marinder}},
  \bibinfo{author}{\bibfnamefont{P.-L.} \bibnamefont{Wang}}, \bibnamefont{and}
  \bibinfo{author}{\bibfnamefont{P.}~\bibnamefont{Werner}},
  \bibinfo{journal}{Acta. Chem. Scand.} \textbf{\bibinfo{volume}{40}},
  \bibinfo{pages}{467} (\bibinfo{year}{1986}).

\bibitem[{\citenamefont{Zhong et~al.}(1994)\citenamefont{Zhong, King-Smith, and
  Vanderbilt}}]{loto}
\bibinfo{author}{\bibfnamefont{W.}~\bibnamefont{Zhong}},
  \bibinfo{author}{\bibfnamefont{R.~D.} \bibnamefont{King-Smith}},
  \bibnamefont{and}
  \bibinfo{author}{\bibfnamefont{D.}~\bibnamefont{Vanderbilt}},
  \bibinfo{journal}{Phys. Rev. Lett.} \textbf{\bibinfo{volume}{72}},
  \bibinfo{pages}{3618} (\bibinfo{year}{1994}),
  \urlprefix\url{http://link.aps.org/doi/10.1103/PhysRevLett.72.3618}.

\bibitem[{\citenamefont{Li et~al.}(2015)\citenamefont{Li, Ren, Guo, and
  He}}]{hyperferro_li}
\bibinfo{author}{\bibfnamefont{P.}~\bibnamefont{Li}},
  \bibinfo{author}{\bibfnamefont{X.}~\bibnamefont{Ren}},
  \bibinfo{author}{\bibfnamefont{G.-C.} \bibnamefont{Guo}}, \bibnamefont{and}
  \bibinfo{author}{\bibfnamefont{L.}~\bibnamefont{He}}, \bibinfo{journal}{arXiv
  preprint arXiv:1510.06835}  (\bibinfo{year}{2015}).

\bibitem[{\citenamefont{Ye}(2016)}]{mengye}
\bibinfo{author}{\bibfnamefont{M.}~\bibnamefont{Ye}},
  \emph{\bibinfo{title}{First-principles study of magnetoelectric effects and
  ferroelectricity in complex oxides}} (\bibinfo{year}{2016}).

\bibitem[{\citenamefont{Castro et~al.}(1994)\citenamefont{Castro, Millan,
  Enjalbert, Snoeck, and Galy}}]{sbwo}
\bibinfo{author}{\bibfnamefont{A.}~\bibnamefont{Castro}},
  \bibinfo{author}{\bibfnamefont{P.}~\bibnamefont{Millan}},
  \bibinfo{author}{\bibfnamefont{R.}~\bibnamefont{Enjalbert}},
  \bibinfo{author}{\bibfnamefont{E.}~\bibnamefont{Snoeck}}, \bibnamefont{and}
  \bibinfo{author}{\bibfnamefont{J.}~\bibnamefont{Galy}},
  \bibinfo{journal}{Materials Research Bulletin} \textbf{\bibinfo{volume}{29}},
  \bibinfo{pages}{871 } (\bibinfo{year}{1994}), ISSN \bibinfo{issn}{0025-5408},
  \urlprefix\url{http://www.sciencedirect.com/science/article/pii/0025540894900078}.

\bibitem[{\citenamefont{Krüger and Müller-Buschbaum}(1992)}]{cubiwo}
\bibinfo{author}{\bibfnamefont{T.}~\bibnamefont{Krüger}} \bibnamefont{and}
  \bibinfo{author}{\bibfnamefont{H.}~\bibnamefont{Müller-Buschbaum}},
  \bibinfo{journal}{Journal of Alloys and Compounds}
  \textbf{\bibinfo{volume}{190}}, \bibinfo{pages}{L1 } (\bibinfo{year}{1992}),
  ISSN \bibinfo{issn}{0925-8388},
  \urlprefix\url{http://www.sciencedirect.com/science/article/pii/0925838892901564}.

\bibitem[{\citenamefont{Perdew and Zunger}(1981)}]{lda1}
\bibinfo{author}{\bibfnamefont{J.~P.} \bibnamefont{Perdew}} \bibnamefont{and}
  \bibinfo{author}{\bibfnamefont{A.}~\bibnamefont{Zunger}},
  \bibinfo{journal}{Phys. Rev. B} \textbf{\bibinfo{volume}{23}},
  \bibinfo{pages}{5048} (\bibinfo{year}{1981}).

\bibitem[{\citenamefont{Ploetz and Mueller~Buschbaum}(1982)}]{pbalo}
\bibinfo{author}{\bibfnamefont{K.}~\bibnamefont{Ploetz}} \bibnamefont{and}
  \bibinfo{author}{\bibfnamefont{H.}~\bibnamefont{Mueller~Buschbaum}},
  \bibinfo{journal}{Zeitschrift fuer Anorganische und Allgemeine Chemie}
  \textbf{\bibinfo{volume}{488}}, \bibinfo{pages}{44} (\bibinfo{year}{1982}).

\bibitem[{\citenamefont{Mahe-Paillert}(1970)}]{vmoo_fe}
\bibinfo{author}{\bibfnamefont{P.}~\bibnamefont{Mahe-Paillert}},
  \bibinfo{journal}{Revue de Chimie Minerale} \textbf{\bibinfo{volume}{7}},
  \bibinfo{pages}{846} (\bibinfo{year}{1970}).

\bibitem[{\citenamefont{Yun et~al.}(2004)\citenamefont{Yun, Ricinschi,
  Kanashima, Noda, and Okuyama}}]{large_polarization}
\bibinfo{author}{\bibfnamefont{K.~Y.} \bibnamefont{Yun}},
  \bibinfo{author}{\bibfnamefont{D.}~\bibnamefont{Ricinschi}},
  \bibinfo{author}{\bibfnamefont{T.}~\bibnamefont{Kanashima}},
  \bibinfo{author}{\bibfnamefont{M.}~\bibnamefont{Noda}}, \bibnamefont{and}
  \bibinfo{author}{\bibfnamefont{M.}~\bibnamefont{Okuyama}},
  \bibinfo{journal}{Japanese Journal of Applied Physics}
  \textbf{\bibinfo{volume}{43}}, \bibinfo{pages}{L647} (\bibinfo{year}{2004}),
  \urlprefix\url{http://stacks.iop.org/1347-4065/43/i=5A/a=L647}.

\bibitem[{\citenamefont{Pailleret et~al.}(1966)\citenamefont{Pailleret, J.,
  Freundlich, and Rimsky}}]{vmoo_af1}
\bibinfo{author}{\bibfnamefont{P.}~\bibnamefont{Pailleret}},
  \bibinfo{author}{\bibfnamefont{B.}~\bibnamefont{J.}},
  \bibinfo{author}{\bibfnamefont{W.}~\bibnamefont{Freundlich}},
  \bibnamefont{and} \bibinfo{author}{\bibfnamefont{A.}~\bibnamefont{Rimsky}},
  \bibinfo{journal}{Comptes Rendus Hebdomadaires des Seances de l'Academie des
  Sciences} \textbf{\bibinfo{volume}{263}}, \bibinfo{pages}{1133}
  (\bibinfo{year}{1966}).

\bibitem[{\citenamefont{Eick and Kihlborg}(1966)}]{vmoo_af2}
\bibinfo{author}{\bibfnamefont{H.}~\bibnamefont{Eick}} \bibnamefont{and}
  \bibinfo{author}{\bibfnamefont{L.}~\bibnamefont{Kihlborg}},
  \bibinfo{journal}{Acta Chemica Scandinavica} \textbf{\bibinfo{volume}{20}},
  \bibinfo{pages}{1666} (\bibinfo{year}{1966}).

\bibitem[{\citenamefont{Galy et~al.}(1999)\citenamefont{Galy, Satto, Sciau, and
  Millet}}]{livo}
\bibinfo{author}{\bibfnamefont{J.}~\bibnamefont{Galy}},
  \bibinfo{author}{\bibfnamefont{C.}~\bibnamefont{Satto}},
  \bibinfo{author}{\bibfnamefont{P.}~\bibnamefont{Sciau}}, \bibnamefont{and}
  \bibinfo{author}{\bibfnamefont{P.}~\bibnamefont{Millet}},
  \bibinfo{journal}{Journal of Solid State Chemistry}
  \textbf{\bibinfo{volume}{146}}, \bibinfo{pages}{129 } (\bibinfo{year}{1999}),
  ISSN \bibinfo{issn}{0022-4596},
  \urlprefix\url{http://www.sciencedirect.com/science/article/pii/S0022459699983184}.

\bibitem[{\citenamefont{Anisimov et~al.}(1991)\citenamefont{Anisimov, Zaanen,
  and Andersen}}]{ldaplusU}
\bibinfo{author}{\bibfnamefont{V.~I.} \bibnamefont{Anisimov}},
  \bibinfo{author}{\bibfnamefont{J.}~\bibnamefont{Zaanen}}, \bibnamefont{and}
  \bibinfo{author}{\bibfnamefont{O.~K.} \bibnamefont{Andersen}},
  \bibinfo{journal}{Phys. Rev. B} \textbf{\bibinfo{volume}{44}},
  \bibinfo{pages}{943} (\bibinfo{year}{1991}).

\bibitem[{\citenamefont{Dudarev et~al.}(1998)\citenamefont{Dudarev, Botton,
  Savrasov, Humphreys, and Sutton}}]{ldaplusU_simplified}
\bibinfo{author}{\bibfnamefont{S.~L.} \bibnamefont{Dudarev}},
  \bibinfo{author}{\bibfnamefont{G.~A.} \bibnamefont{Botton}},
  \bibinfo{author}{\bibfnamefont{S.~Y.} \bibnamefont{Savrasov}},
  \bibinfo{author}{\bibfnamefont{C.~J.} \bibnamefont{Humphreys}},
  \bibnamefont{and} \bibinfo{author}{\bibfnamefont{A.~P.}
  \bibnamefont{Sutton}}, \bibinfo{journal}{Phys. Rev. B}
  \textbf{\bibinfo{volume}{57}}, \bibinfo{pages}{1505} (\bibinfo{year}{1998}).

\bibitem[{\citenamefont{Cococcioni and de~Gironcoli}(2005)}]{dftU_simp}
\bibinfo{author}{\bibfnamefont{M.}~\bibnamefont{Cococcioni}} \bibnamefont{and}
  \bibinfo{author}{\bibfnamefont{S.}~\bibnamefont{de~Gironcoli}},
  \bibinfo{journal}{Phys. Rev. B} \textbf{\bibinfo{volume}{71}},
  \bibinfo{pages}{035105} (\bibinfo{year}{2005}),
  \urlprefix\url{http://link.aps.org/doi/10.1103/PhysRevB.71.035105}.

\end{thebibliography}

\end{document}